\documentclass[aip,jcp,twocolumn,showpacs,nobibnotes,reprint]{revtex4-1}
\usepackage{graphics,graphicx,amsfonts,amsmath,amsbsy,amssymb,color}
\usepackage{verbatim}  
\usepackage{psfrag}  
\usepackage{tabularx}
\usepackage{xmpmulti}
\usepackage{hyperref}
\usepackage{marginnote}
\usepackage{subfigure}
\usepackage{bibentry}
\usepackage{siunitx}
\usepackage{paracol}
\usepackage{xparse}
\usepackage{layouts}
\usepackage{afterpage}
\usepackage{braket}
\usepackage{paralist}
\usepackage{bm}
\usepackage{url}
\usepackage[normalem]{ulem}
\usepackage{color}
\usepackage[colorinlistoftodos]{todonotes}
\usepackage[american]{babel} 

\newcommand{\refeq}[1]{{Eq.~(\ref{#1})}}
\newcommand{\reffig}[1]{{Fig.~\ref{#1}}}
\newcommand{\refsec}[1]{{Sec.~\ref{#1}}}

\newcommand{\kgrid}[1]{{${#1}\times{#1}\times{#1}$}}

\hypersetup{breaklinks,colorlinks=true,linkcolor=blue,citecolor=blue,urlcolor=blue,filecolor=blue}

\begin{document}
\title{Machine learning for a finite size correction in periodic coupled cluster theory calculations}
\author{Laura~Weiler}
\affiliation{Department of Chemistry, University of Iowa}
\author{Tina~N.~Mihm}
\affiliation{Department of Chemistry, University of Iowa}
\author{James~J.~Shepherd}
\email{james-shepherd@uiowa.edu}
\affiliation{Department of Chemistry, University of Iowa}

\date{\today}

\begin{abstract}
We introduce a straightforward Gaussian process regression (GPR) model for the transition structure factor of  metal periodic coupled cluster singles and doubles (CCSD) calculations.
This is inspired by the method introduced by Liao and Gr\"uneis for interpolating over the transition structure factor to obtain a finite size correction for CCSD [J. Chem. Phys. 145, 141102 (2016)], and by our own prior work using the transition structure factor to efficiently converge CCSD for metals to the thermodynamic limit [Nat. Comput. Sci. 1, 801 (2021)]. 
In our CCSD-FS-GPR method to correct for finite size errors, we fit the structure factor to a 1D function in the momentum transfer, $G$.
We then integrate over this function by projecting it onto a k-point mesh to obtain comparisons with extrapolated results. 
Results are shown for lithium, sodium, and the uniform electron gas.
\end{abstract}

\maketitle

\section{Introduction}

Calculations at the thermodynamic limit (TDL) are necessary for understanding properties of solids.
Wavefunction-based methods such as coupled cluster theory are popular for treating electron correlation in many-body solid-state systems. 
However, TDL calculations for periodic solids suffer from slow convergence with increased system size, especially at the coupled cluster level of theory.
Due to the scaling of the coupled cluster method, energies are often calculated at reduced system sizes and extrapolated to the TDL.
Calculating energies at reduced system sizes introduces finite size effects: errors resulting from the modeling of a real solid with a finite number of periodic cells.
Despite this challenge, coupled cluster with single and double excitations (CCSD) has previously been applied to three-dimensional atomistic solids, \cite{gruber_applying_2018, liao_communication:_2016} 
solids in metallic, insulating, and semiconducting phases, \cite{mihm2021shortcut}
and rare-gas solids, \cite{rosciszewski_ab_2000}
and used as a benchmark for comparison against density functional theory (DFT), \cite{kiohara_dft_2013} full configuration interaction quantum Monte Carlo (FCIQMC), \cite{booth_towards_2013} and explicitly correlated second-order perturbation theory. \cite{gruneis_efficient_2015}
Work has also been done to develop equation-of-motion \cite{lewis_ab_2019, wang_absorption_2021, wang_excitons_2020, mcclain_gaussian-based_2017}
and time-dependent \cite{white_time-dependent_2018} 
coupled cluster formalisms for applications to solids and warm dense matter.
Efforts to extend coupled cluster to solids have benefited from the various finite size corrections developed by the quantum Monte Carlo (QMC) community. \cite{fraser_finite-size_1996, holzmann2016theory, drummond_finite-size_2008, azadi2015systematic}
These developments include extrapolation methods,
\cite{holzmann_momentum_2011, ruggeri_correlation_2018} 
the identification of a special twist angle, \cite{rajagopal_quantum_1994, dagrada_exact_2016}
the replacement of the Coulomb potential with Ewald interactions, \cite{williamson_elimination_1997, fraser_finite-size_1996}
and corrections based on the random phase approximation (RPA) \cite{chiesa_finite-size_2006} and the KZK functional. \cite{kwee_finite-size_2008}
Twist averaging has also been developed for addressing finite size effects in
variational, \cite{lin_twist-averaged_2001} 
diffusion, \cite{drummond_finite-size_2008} 
and full configuration interaction Monte Carlo methods. \cite{ruggeri_correlation_2018}
In addition to developments in finite size corrections, 
embedding methods \cite{sun_quantum_2016} 
aim to alleviate finite size error by injecting high-level theory into finite regions of systems where an overall high-level treatment is infeasible.
CCSD may be used to attain accurate descriptions of correlation in finite regions of systems without incurring the full overhead of a general CCSD calculations.
To this aim, CCSD has recently been embedded in RPA. \cite{schafer_local_2021}
Despite advantages of embedding theories, finite size corrections are still necessary for applications where an overall high-level treatment is desired.

Liao and Gr\"uneis recently introduced a finite size correction for periodic coupled cluster calculations (CCSD-FS) based on the transition structure factor. \cite{liao_communication:_2016}
Their CCSD-FS theory has demonstrated a reduction in finite size error for insulating solids such that energies are converged to within chemical accuracy with \kgrid{3} k--meshes. %
CCSD-FS has since been used in applications such as the adsorption of water molecules on boron nitride, \cite{gruber_applying_2018}
and for studying the electronic properties of solid hydrogen crystals. \cite{liao2021structural}
However, only insulators were included in the original work by Liao and Gr\"uneis due to the exacerbation of finite size effects in small and zero gap systems.
In our recent work locating a special twist angle to converge CCSD correlation energies for metals, \cite{mihm2021shortcut}
we applied CCSD-FS to small gap systems but our application required an additional extrapolation to fully remove finite size error.
This paper is intended as a direct extension of the CCSD-FS theory.
In part inspired by recent machine learning approaches across the chemical sciences,
we seek to
determine whether structure factor interpolation via Gaussian process regression (GPR) can be used to provide a CCSD-FS-level correction for metals.
We benefit from the GPR approach not requiring an explicit parametrization of the transition structure factor. 
For this proof of principle work we took the ground truth for GPR to be the structure factor of the largest system size we could calculate. Then energies were calculated by, first, projecting the resulting continuous function onto auxiliary grids of varying sizes up to this maximum grid size, then, second, extrapolating these to an infinite grid size. These CCSD-FS-GPR energies were then compared with CCSD, extrapolated CCSD over the same grid sizes, CCSD-FS, and CCSD-FS with additional extrapolation.

\section{CCSD and the transition structure factor}\label{sec:CCSD for solids}

We employ the coupled cluster method in the typical fashion as described in previous studies. \cite{shepherd_many-body_2013,shepherd_range-separated_2014,shepherd_coupled_2014}
Here, we will only provide a brief summary overview of the method.

Coupled cluster expresses the ground-state wavefunction using an exponential operator, $e^{\hat{T}}$, acting on the reference Hartree--Fock wavefunction. 
The coupled cluster energy is typically expressed as a sum of the Hartree--Fock energy plus the coupled cluster correlation energy correction calculated using the following equation: 
\begin{equation}
 E_\mathrm{corr} = \frac{1}{2}\sum_{ijab} T_{ijab} V_{ijab}
\label{Corr_E}  
\end{equation}
Here, $T_{ijab}$ are the individual t-amplitudes for the excitation operator, $\hat{T}$, and $V_ {ijab}$ is the antisymmeterized electron repulsion integral defined as $V_ {ijab} = 2v_{ijab} – v_{ijba}$.
In this paper, we only work with the coupled cluster singles and doubles (CCSD) energy, so $ T_{ijab}$ is therefore defined as $T_{ijab} = t_{ijab} + t_{ia}t_{jb}$, where $t_{ia}$ and $t_{ijab}$ correspond to the singles and doubles amplitudes, respectively.
Note that for the UEG, singles amplitudes are zero by momentum symmetry and CCSD therefore reduces to CCD.
For all equations, $i$ and $j$ refer to occupied orbital indices and $a$ and $b$ refer to unoccupied orbital indices in the wavefunction. 

Following the derivation laid out in the paper by Liao and Gr\"uneis, \cite{liao_communication:_2016} the correlation energy can be re-written in terms of the transition structure factor, $S_{\bf G}$:
\begin{equation}
 E_\mathrm{corr} = \frac{1}{2}\sum'_{\bf G} S_{\bf G} V_{\bf G}
\label{SF_Corr_E}  
\end{equation}
where $\bf G$ is the plane wave vector that is a sum of the reciprocal space lattice vector, $\bf g$, and the difference between two crystal lattice vectors chosen within the first Brillouin zone, $\Delta {\bf k}$. Here $V_{\bf G}$ is the electron repulsion integral as a function of ${\bf G}$, taking the form $4\pi/G^2$ (where $G$ is the magnitude of ${\bf G}$). %

Crucially, in order to correct for finite size errors, we wish to recover the CCSD correlation energy \emph{not captured} by the sum in \refeq{SF_Corr_E}. The missing energy comes from the discretization of ${\bf G}$, leading to quadrature error, and the missing component of the integral between some minimum $|{\bf G}|$ and $|{\bf G}|=0$:
$\int_{|{\bf G}|=0}^{|{\bf G}|=G_\mathrm{min}} S_G V_G \, dG$. Here, $G_\mathrm{min}$  is the minimum non-zero value of G present in the calculation grid.

\section{Connectivity twist averaging}
To help reduce finite size effects (FSE) in general, we used our connectivity twist averaging (cTA) method from previous work. \cite{mihm_optimized_2019} 
This method is based on twist averaging, which reduces finite size effects through reducing the fluctuations in the wavefunction. This is done through using offsets (also called twist angles) which are applied to the grid points such that ($\phi_j \propto \exp(i ({\bf k}_j - {\bf k}_s)\cdot {\bf r})$), where $k_s$ is the offset.  
Typically, $N_s$ twist angles are run, and the resulting energies across twist angles are then averaged over to obtain the final twist averaged energy, such that: 
\begin{equation*}
    \langle E_\mathrm{corr} \rangle _{k_s}= \frac{1}{N_s} \sum_{l=1}^{N_s} E_\mathrm{corr}(k_{s,l})
\end{equation*}
This results in a linear cost scaling factor of $N_s$ for twist averaging. In contrast, cTA aims to reduce this linear cost by finding a single twist angle that reproduces the twist averaged energy. This is done through looking at the momentum transfer vectors between the occupied ($i$) and virtual ($a$) orbital manifolds, 
$x=|{\bf n_i}-{\bf n_a}|^2$,
Here, ${\bf n_i}$ is the integer equivalent of the quantum number associated with momentum: 
${\bf k_i}=\frac{2\pi}{L} {\bf n_i}$.
For each twist angle, these transfers are then histogrammed over the same indices as the correlation energy in \refeq{Corr_E} to form a vector with elements $h_x$. 
This vector measures what we call the connectivity. This histogramming is equivalent in cost to an MP2-like calculation and is repeated for 100 randomly chosen twist angles
The histogramming vector is then averaged over all twist angles to give $\langle {\bf h}\rangle_{k_s}$, which is then used to find the twist angle with the minimum residual difference to the average using the following relationship:
\begin{equation*}
    S_\mathrm{res}({\bf k_s})=\sum_x \frac{1}{x^2}|h_x({\bf k}_s)-\langle h_x\rangle_{k_s}|^2
\end{equation*}
Here, $S_\mathrm{res}({\bf k_s})$ is the residual difference to the average for a single twist angle and $\frac{1}{x^2}$ weights towards the shorter momentum transfers as they contribute more to the energy. The ${\bf k_s}$ found with the smallest residual difference to the averaged connectivity is the cTA special twist angle, ${\bf k_s}^*$, which we have found can reproduce the twist average energy and, therefore, has the smallest amount of FSE in the energy compared to the other twist angles. With this special twist angle, we can then run a single coupled cluster calculation using \refeq{SF_Corr_E}, such that:
\begin{equation*}
    E_{corr}({\bf k_s}^*)=\frac{1}{2}\sum'_{G} S_G(k_{s}^{*})V_G(k_{s}^{*})
\end{equation*}
As this method depends on the momentum transfers that occur in a given system, a different special twist angle is found for each system.

It is important to note here that our cTA method has only been developed for use in the uniform electron gas so, while all UEG calculations shown were run directly with our cTA method to get the structure factors and energies used in the GPR fits, we had to use another strategy for lithium and sodium. For Li and Na, we used a UEG calculation to find the cTA special twist angle before running a CCSD calculation using the realistic Hamiltonian to obtain the transition structure factors and energies used in the GPR fits. The parameters for the UEG calculations were set to have the same number of electrons as the supercell equivalent to the k-point grid (e.g. for a \kgrid{2} grid we have 16 electrons). We note that the density does not affect the twist angle chosen, and so the same special twist angle (per k-point grid) was used for both the Li and Na systems.

\begin{figure}
\subfigure[\mbox{}]{%
\includegraphics[width=0.5\textwidth,height=\textheight,keepaspectratio]{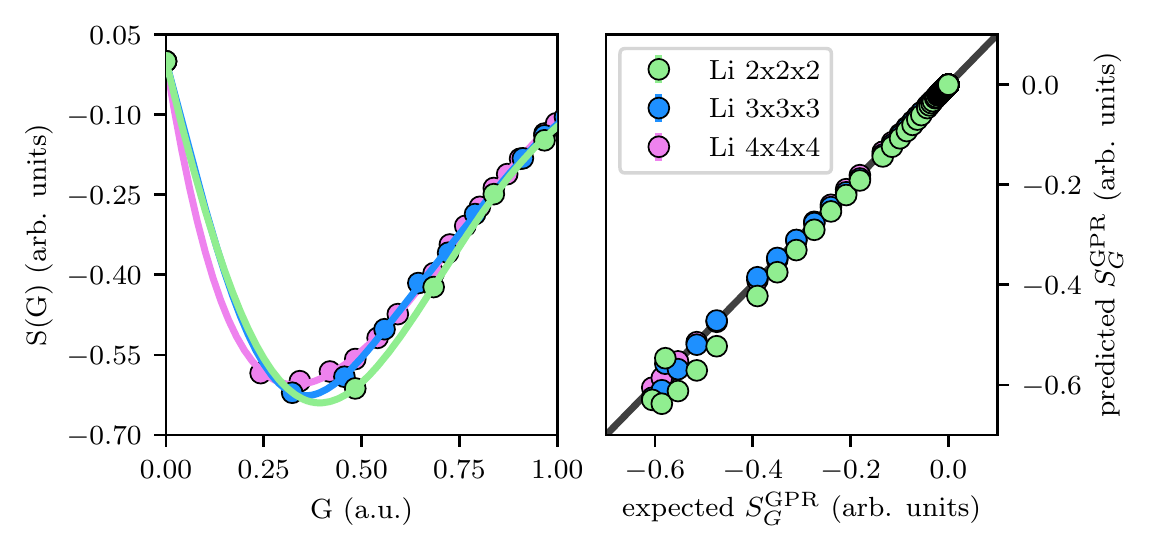}
\label{subfig:Li-fits}
}
\subfigure[\mbox{}]{%
\includegraphics[width=0.5\textwidth,height=\textheight,keepaspectratio]{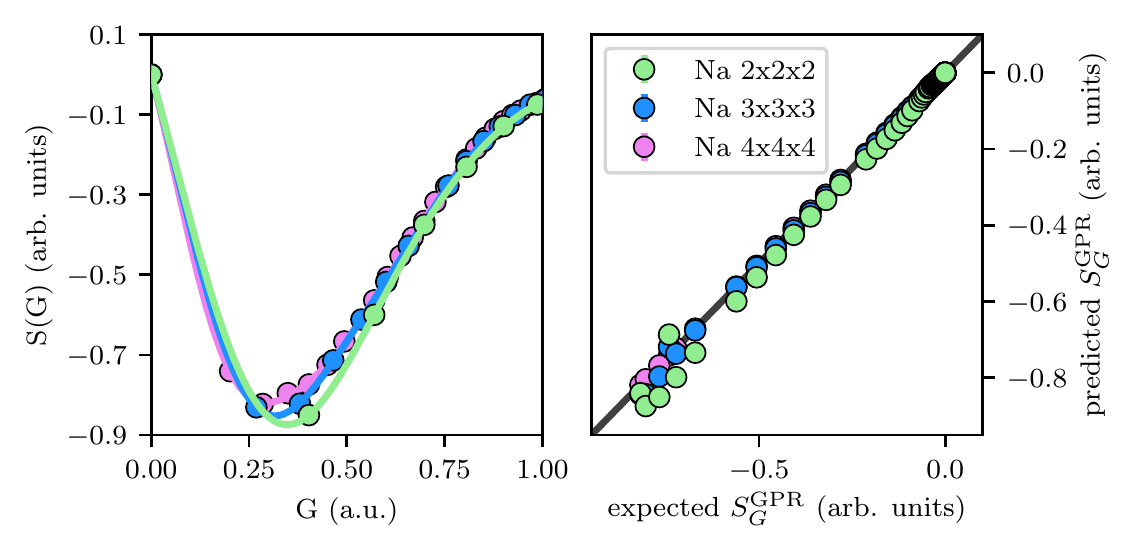}
\label{subfig:Na-fits}
}
\subfigure[\mbox{}]{%
\includegraphics[width=0.5\textwidth,height=\textheight,keepaspectratio]{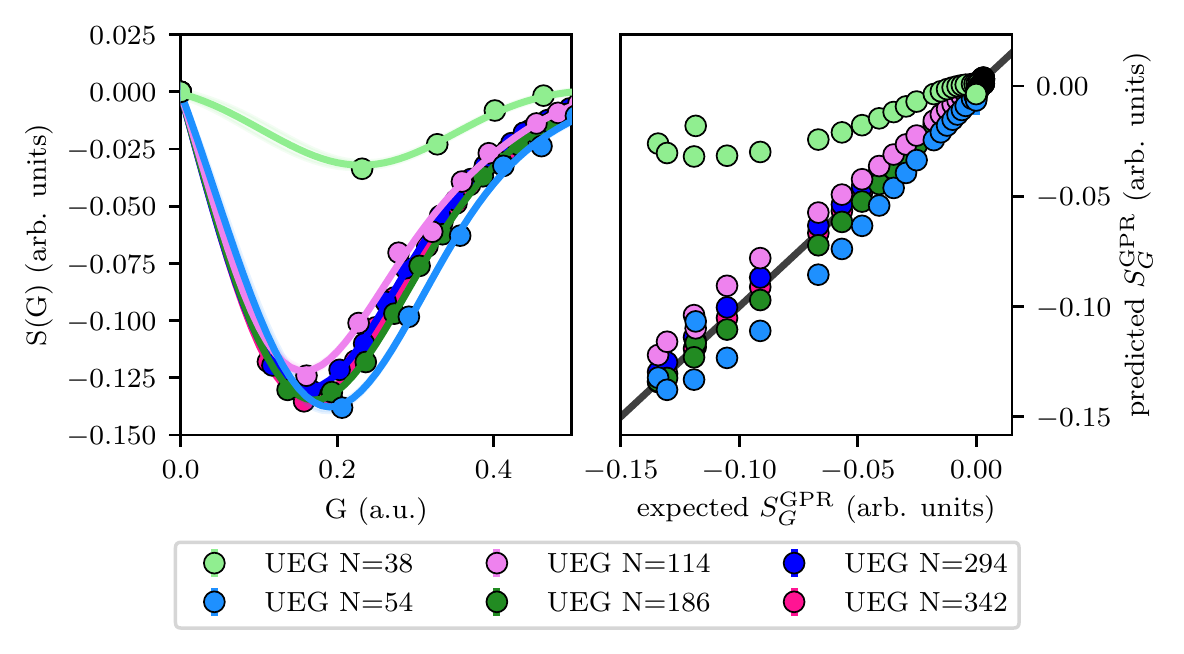}
\label{subfig:UEG-fits}
}
\caption{
Data from the GPR approach to fitting the transition structure factor for: (a) bcc Li, (b) bcc Na, and (c) UEG $r_S=5$ a.u. In each panel we show GPR transition structure factor fits for a range of grid sizes on the left, and on the right we compare the transition structure factor for the largest grid (expected) against the GPR fits for each grid (predicted) where an $x=y$ line represents perfect agreement between the expected and predicted transition structure factors. Note that $2\sigma$ confidence intervals are plotted on each of the GPR fits but are too small to see.}
\label{fig:fits}
\end{figure}

\section{GPR protocol}\label{sec:protocol}

Gaussian process regression makes use of the ``kernel trick'' \cite{williams2006gaussian} to flexibly interpolate data for a wealth of applications.
There has been a large amount of recent work using GPR for molecular and materials chemistry \cite{deringer2021gaussian} 
including the use of GPR to predict molecular crystal energies, \cite{krynski2021efficient, raimbault2019using}
MP2 and CCSD energies, \cite{welborn_transferability_2018}
potential energy surfaces, \cite{kamath2018neural}
exchange spin coupling, \cite{bahlke2020exchange}
and for autonomous materials discovery. \cite{noack2020autonomous}
Recently, GPR was used in a finite size correction for the coupled cluster and auxiliary field quantum Monte Carlo energies of one-dimensional hydrogen
chains. \cite{borda2021gaussian}

One reason for GPRs growth in popularity is that it is nonparametric, or requires no \emph{a priori} assumption about the distribution of data. 
For our application this quality is beneficial because we do not need to assume a functional form for the fit.
GPR models can also be successfully trained on small data sets, making them a relatively efficient surrogate for many-body methods as GPR scales as $\mathcal{O}(n^3)$ with input size $n$.

We use GPR because it is well-suited to our design constraints, such as the provision of a smooth, continuous, one-dimensional function which is integrable on the $G$ domain.
We tested both GPR and neural networks for fitting transition structure factors and found the GPR approach to be less prone to overfitting. We also tried multiple frameworks for GPR in python, and settled on the open-source {\tt scikit-learn} \cite{scikit-learn} library for its simple user interface.

Our GPR models require the specification of a kernel and an alpha (or noise) value.
The GPR approach is known to be sensitive to the choice of kernel. \cite{williams2006gaussian} %
For our kernel specification we sought a single kernel which would provide reasonable fits across systems for simplicity.
Prior to this design choice, we tried a kernel selection via grid search but found this impractical due to a lack of validation metric which accounts for the physics of the transition structure factor.
In practice, we found that a sum of two radial basis function (RBF) kernels provided enough flexibility to fit all of our systems.
Our kernel is initially set with a length scale of 0.5 and a variance of 1.0, and then tuned to optimal values by maximizing the log-marginal likelihood during training.
We use alpha values of $10^{-5}$ to $10^{-6}$ to account for noise in the transition structure factor. 
We train our GPR model on $G$, $\overline{S_G}$ pairs in order to learn a continuous, 1D representation of the transition structure factor, $S(G)$, which we can then project onto discrete grids for use in \refeq{SF_Corr_E} (see the Supplementary Material for details concerning the transformation from  $S_{\bf G}$ to $\overline{S_G}$).
To generate structure factor predictions, we perform a weighted average over ten trained models' output using a weighting of $1/\sigma^2$, where $\sigma$ is one standard error returned with each prediction from the {\tt scikit-learn} {\tt predict} method when the {\tt return\_std} flag is set to {\tt True}.
This averaging scheme rewards large confidence in predictions, and penalizes predictions with large uncertainty. By averaging over multiple models we additionally increase confidence in predictions and encourage reproducibility without having to seed the random number generator in the L-BFGS-B optimizer used by {\tt scikit-learn} to maximize the log marginal likelihood.

We train a GPR model for each Li, Na, and UEG grid size and then use each trained model to predict transition structure factors on all grid sizes (see \refsec{sec:calculations} for details on the systems used). 
We treat the largest grid size for each system as the ground truth for assessing the performance of our models after training (see the right panels of \reffig{subfig:Li-fits}, \reffig{subfig:Na-fits}, and \reffig{subfig:UEG-fits} for a visual depiction and Section \refsec{sec:results} for discussion).
We then use the GPR-predicted transition structure factors to calculate correlation energies at each auxiliary grid size by taking the sum in \refeq{SF_Corr_E}.
We then extrapolate the prediction from each model across different grid sizes to the infinite grid limit to calculate the CCSD-FS-GPR energy. The technical details for this are in the Supplementary Material.

\section{Calculations}\label{sec:calculations}
We calculated coupled cluster transition structure factors and correlation energies for the lithium and sodium solids.
The UEG systems were parameterized by a Wigner-Seitz radius of $r_s=5$~a.u. and system sizes from $N=38$ electrons to $342$ electrons. The bcc Li and bcc Na systems were parameterized by k-point mesh grids from \kgrid{2} to \kgrid{4} (equivalently, $N=16$ to $128$), and lattice constants $a=3.363 \textup{~\AA}$ and $a=4.047 \textup{~\AA}$, respectively. 
Lattice constants were taken from the NOMAD database. \cite{scheffler_nomad_2015}
The UEG calculations are from an in-group uegccd code,\cite{shepherd_range-separated_2014,shepherd_coupled_2014} and Li and Na calculations are from the Vienna Ab initio Simulation Package (VASP).\cite{kresse_norm-conserving_1994, kresse_efficient_1996}
Arbitrary units (denoted arb.~units) are used for the transition structure factor and Hartree atomic units are used for $G$ and $r_s$ values.

\section{Results}\label{sec:results}

A summary of our protocol is: 1) to fit the continuous function $S(G)$ using the GPR procedure outlined in \refsec{sec:protocol}, 2) to compare these to transition structure factors at the largest system size (which is our best known data), and 3) to use these fits to estimate a TDL energy and compare with known results. 
For Li, CCSD-FS is the state-of-the-art and our benchmark for comparison, these data were found to require additional extrapolation to reach the TDL by a previous study. \cite{mihm2021shortcut}
The benchmarks used for comparison for Na and the UEG are extrapolated CCSD and CCD, respectively. 
Of these benchmarks, the Li and UEG numbers are likely the closest to true TDL benchmarks – from CCSD-FS plus extrapolation and extrapolation over large system sizes, respectively. Na is benchmarked against extrapolated CCSD without the FS correction and is therefore less likely to be at the TDL.

In \reffig{subfig:Li-fits} we show a complete set of GPR fits for Li.
The left panel of \reffig{subfig:Li-fits} shows transition structure factors from the \kgrid{2}, \kgrid{3}, and \kgrid{4} k-meshes. The number of data points these have, respectively, is: 41, 90, and 159.
GPR fits these curves well and a typical deviation between the GPR line and the points is 0.0004 at each grid size.  %

The prediction of the point at the origin is particularly important for this method as we know that the transition structure factor must fall to zero.
The predictions are: $-0.0008(10)$ for \kgrid{2} and \kgrid{3}, and $-0.0015(10)$ for \kgrid{4}. The numbers in parentheses are the error on each value. 
In general, these values seem reasonable.
It is noteworthy that there is a slight, visible change in the position of the structure factor minimum as the grid is increased in size; this is likely due to relaxation of the coupled cluster $t$ amplitudes in the presence of the increased number of basis functions. 
This kind of minimum drifting can make finite size effects more severe. 

\begin{figure}
\subfigure[\mbox{}]{%
\includegraphics[width=0.44\textwidth,height=\textheight,keepaspectratio]{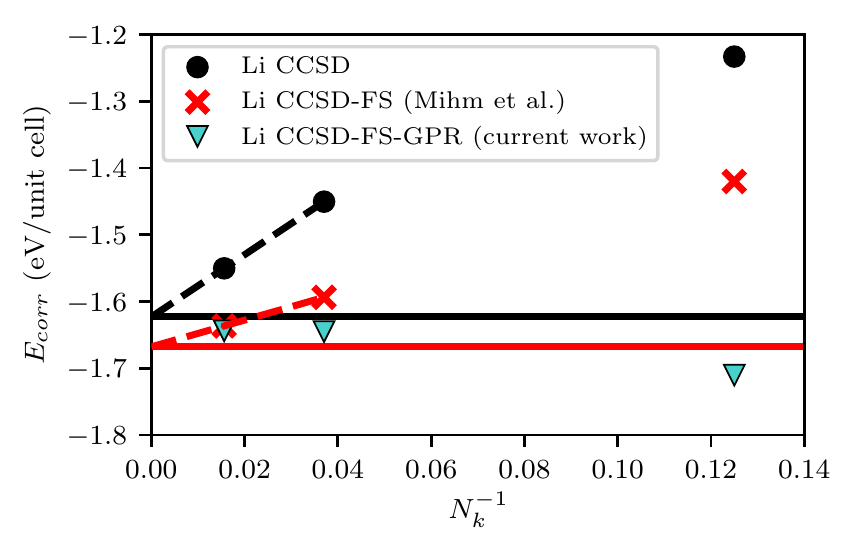}
\label{subfig:Li-energies}
}
\subfigure[\mbox{}]{%
\includegraphics[width=0.44\textwidth,height=\textheight,keepaspectratio]{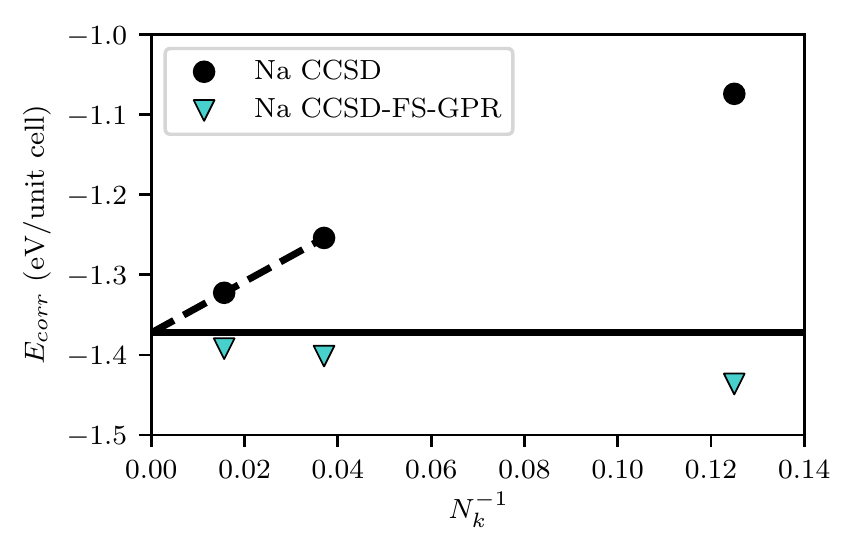}
\label{subfig:Na-energies}
}
\subfigure[\mbox{}]{%
\includegraphics[width=0.46\textwidth,height=\textheight,keepaspectratio]{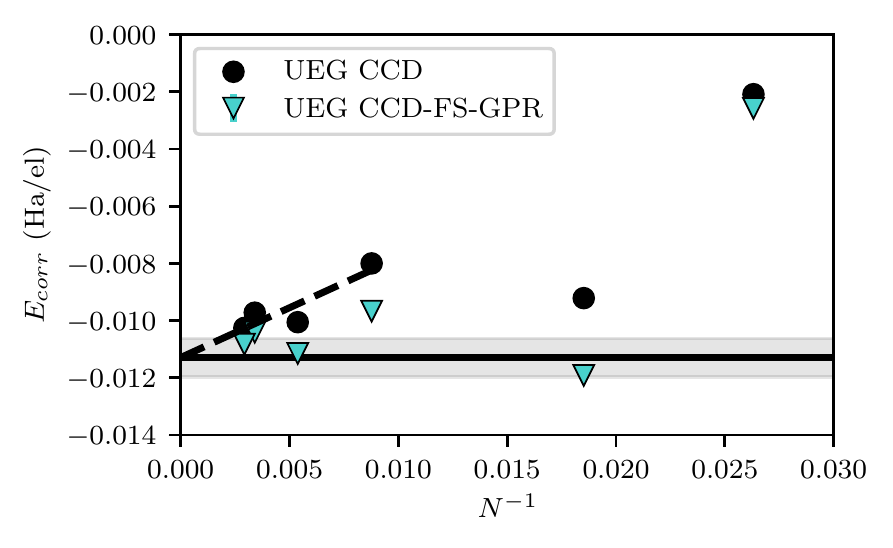}
\label{subfig:UEG-energies}
}
\caption{
CCSD-FS-GPR energy results compared to CCSD for: (a) bcc Li, (b) bcc Na, and (c) UEG $r_S=5$~a.u. 
Horizontal lines show results extrapolated to the thermodynamic limit, with dashed lines showing the points used and function fit for the extrapolation. In panel (a), Mihm et al. CCSD-FS benchmark data comes from Ref.~\onlinecite{mihm2021shortcut} and these authors perform the extrapolation shown here in red (for more discussion, see main text). In panel (c), the shaded region corresponds to plus/minus one standard error on the 4-point CCD extrapolation; the remaining extrapolations in panels (a) and (b) are 2-point extrapolations and therefore do not have a standard error.
}
\label{fig:energies}
\end{figure}

The right panel of figure \reffig{subfig:Li-fits} shows these same bcc Li transition structure factors with the GPR function evaluated at grid points corresponding to \kgrid{4}. 
The GPR predictions are plotted against the expected (ground truth) $S_G$ which is taken as the \kgrid{4} transition structure factor. %
The high $G$ tail of the transition structure factor makes up the vast majority (97.4$\%$, 96.7$\%$, and 96.9$\%$ in order of increasing grid size) of the points, and the function is well-fit in this region of the curve (i.e. within 0.049, 0.012, 0.007 of the ground truth mean).
While not the focus of this paper, this is important because the tail is responsible for basis set convergence. \cite{liao_communication_2016}
Its lack of change as the system size changes (either in the GPR or in the presumed $t$ amplitudes) lends credence to the approach of separating out the basis set and finite size errors because they commute.\cite{shepherd_communication:_2016}
At larger transition structure factor values, $|S_G| > 0.5$, we see that there are discrepancies between what GPR on \kgrid{2} data predicts and what is found at \kgrid{4}.
Owing to the low fit error in GPR (0.0004) relative to the scale of the transition structure factors, %
we attribute this to the effects of the structure factor minimum drifting described above rather than to the fitting procedure.

The GPR protocol is repeated for Na and the UEG in the remainder of \reffig{fig:fits}.
The transition structure factor fits for both systems (\reffig{subfig:Na-fits} and \reffig{subfig:UEG-fits})
are comparable with Li, except that Na shows slightly less relaxation with increasing system size. 
The opposite tendency is seen in the UEG.
In the right panel of \reffig{subfig:Na-fits} the average deviation between the GPR-predicted Na \kgrid{4} point and the line is comparable to the Li result: 0.022, 0.008, and 0.004 in order of increasing grid size. The deviation is slightly larger for the UEG in \reffig{subfig:UEG-fits}, ranging from 0.003 at large $N$ to 0.109 at small $N$. Both Na and UEG predictions at the $G=0$ point are reasonable at all grid sizes and comparable to those of Li.

To get a better sense of these GPR fit errors, we can convert them to energies by comparing the CCSD energy of a given k-point grid with an energy evaluation of the GPR-trained $S_G$ projected onto the same grid. We find that these energies are generally small compared to the overall energy. For Li and Na they are $<3$~meV/unit cell and $<0.4$~meV/unit cell respectively; for the UEG they are $<3$~mHa/el decreasing to $<1$~mHa/el at $N>186$. A full list of these values is presented in the supplementary material.

Figure~\ref{fig:energies} compares CCSD-FS-GPR results with extrapolated CCSD. %
In \reffig{subfig:Li-energies} bcc Li TDL energy estimates from GPR predictions at the three grid sizes are compared with CCSD-FS literature values from our recent paper.\cite{mihm2021shortcut}
In that paper, the CCSD-FS method was used on a different twist angle selection protocol analogous to our cTA method here (see \refsec{sec:CCSD for solids}).
The CCSD-FS data were then extrapolated to the TDL using $1/N$ owing to a remaining error in the functional form of the fit. \cite{mihm2021shortcut}
This is our best estimate for the TDL of this system. 
We also include another power-law fit from conventional $1/N$ extrapolation to our CCSD data. 

All three of our bcc Li CCSD-FS-GPR energies are considerably improved over the raw CCSD correlation energies for each grid size. 
The CCSD-FS-GPR energies for \kgrid{3} and \kgrid{4} lie between extrapolated CCSD and CCSD-FS (with additional extrapolation).
We see this as consistent with the need to extrapolate after CCSD-FS corrections in metals.
We also see that the GPR correction for \kgrid{2} is almost twice as large as that of CCSD-FS. 
This trend in how CCSD-FS-GPR behaves for Li matches the behavior of CCSD-FS in insulators and improves over CCSD-FS without extrapolation in metals.

CCSD-FS-GPR results are presented for Na and the UEG in the remainder of \reffig{fig:energies}.
As with Li, all three of the bcc Na CCSD-FS-GPR predictions improve over both the raw CCSD results. %
In \reffig{subfig:Na-energies}, the \kgrid{3} and \kgrid{4} Na grid sizes predict similar energies of $-1.402$~eV/unit cell and $-1.392$~eV/unit cell, respectively, which falls slightly lower than the extrapolated CCSD number ($-1.372$~eV/unit cell). A downward trend in the TDL predictions is visible across the three grids, with \kgrid{2} predicting a lower value of $-1.436$~eV/unit cell.
We attribute this to a similar effect seen in \reffig{subfig:Li-energies} for Li and in general believe the \kgrid{3} and \kgrid{4} predictions to be reasonable and the \kgrid{2} prediction to be greatly improved over the CCSD number.

In \reffig{subfig:UEG-energies}, we show CCSD-FS-GPR results for the UEG.
The UEG CCSD-FS-GPR result converges nicely to the extrapolated CCSD with increased system size and agrees with the extrapolated CCSD within error for grid sizes $N=54$, $186$, and $342$.
Unlike for Na and Li, the smallest grid size ($N=38$) barely improves over the raw CCSD energy. This is likely due to the points in $N=38$ being severely different than the other system sizes 
(see \reffig{subfig:UEG-fits}).
However, the CCSD-FS-GPR energies generally improve over the raw CCSD correlation energies %
and the discrepancies between predictions at different grid sizes is likely due to the structure factor minimum drifting which is most severe for the UEG.

\section{Conclusions}

In this paper we provide a straightforward application of Gaussian process regression (GPR) for correcting finite size effects in small metallic systems. This application is an extension of work previously done by Liao and Gr\"uneis on transition structure factor interpolation for correcting finite size errors in coupled cluster calculations for insulators. \cite{liao_communication:_2016} 
We have demonstrated in this application that transition structure factor interpolation via GPR can correct finite size error in simple metals to achieve CCSD-FS level accuracy; we call the resulting method CCSD-FS-GPR.

We note that we did not attempt to address the accuracy of CCSD in this manuscript. CCSD was our model of choice for finite size effects because of a trade off between accuracy and cost. However, it is also believed that CCSD will capture the physical effects of what is happening on approach to the thermodynamic limit (i.e. the long-range effects), in part because it contains the same physical information as the random phase approximation (RPA).
This paper highlights the importance of the minimum of the structure factor, which appears to change with system size (and presumably method). Our initial investigations have found this to be a complicated phenomenon and we will save its discussion for a separate study.

We believe these results will generalize to different electronic structures, atomic compositions, and lattice symmetries; though there are formalism challenges related to including anisotropy that will need to be overcome. This is where our attention will focus in our next study.

As previously referenced, a similar GPR approach for finite size correction in solids was very recently published by Borda and Rubenstein as a preprint. \cite{borda2021gaussian}
While the work of Borda and Rubenstein is similar to our approach in the motivation for using the GPR algorithm, a fundamental difference in the two approaches is the subject of the regression. Where we learn a representation of the CCSD transition structure factor which is then summed to attain CCSD correlation energies, Borda and Rubenstein learn correlation energies from GPR models trained on smooth overlap of atomic positions (SOAP) descriptors. We believe that our work is consistent with the evidence provided by Borda and Rubenstein that GPR is a good tool for finite size correction in costly many-electron calculations.

\section{Supplementary Material}
The reader is directed to the Supplementary Material for details concerning extrapolation methods.

\section{Acknowledgements}

The research presented here was funded by the National Science Foundation under NSF CHE-2045046.
The University of Iowa is also acknowledged for computer time.
For the purposes of providing information about the calculations used, files will be deposited with Iowa Research Online (IRO) with a reference number [to be inserted].
The modifications of the VASP code used to calculate the transition structure factor and CCSD energies were made in a local repository based on VASP 5.4/VASP 6, which is intended for future release.
VASP and its source code is software available for purchase.
We would like to thank Kevin Robben for discussion and encouragement. 

\section{Data Availability}

The data that support the findings of this study are available within the article.

 \end{document}